# Ultrafast hard X-ray sources based on Relativistic Electrons at ELI Beamlines

U. Chaulagain[1], M. Lamač[1], M. Raclavský[1], S. Karatodorov[1,†], J. Nejdl[1], and S. V. Bulanov[1]

[1]*ELI Beamlines, Institute of Physics ASCR, Prague, Czech Republic*

*†Present affiliation: Institute of Solid-State Physics Bulgarian Academy of Sciences, Bulgaria*

## 1. Introduction

One of the main objectives of ELI beamlines is to deliver ultrashort, bright beams of short wavelength radiation, both coherent and incoherent, to the broader user community [1,2]. Applications of these ultrafast X-ray sources range from structure analysis in atomic physics and molecular chemistry, in solid-state physics, as well as in plasma physics, and fundamental research. The ultrashort nature of these sources provides imaging of a sample with atomic resolution both in space and time. ELI beamlines implements different beamlines providing a unique combination of X-ray sources to the broader user community. A dedicated coherent XUV radiation source with femtosecond pulses from the HHG beamline has been operational in the experimental hall E1 for user experiments [3]. The Gammatron beamline [4] provides incoherent X-ray sources of radiation with tunable photon energy ranging from few keV to MeV. A separate betatron radiation source has also been developed in E3 for Plasma physics research [5].

## 2. Gammatron beamlines

The Gammatron beamline, located at the experimental hall E2, is a multidisciplinary user-oriented source of hard X-ray radiation based on the laser-plasma accelerator (LPA). The beamline provides X-rays with tunable photon energy with a spectral range from a few keV to few hundreds of keV through a betatron scheme and a quasi-monoenergetic beam of hard X-rays up to MeV in inverse Compton scheme. The mechanism of generating such X-ray pulses is laser wakefield acceleration in which an intense ultrashort laser beam propagates through gas-jet targets to induce a strong plasma wake, which accelerates electrons into a well-collimated beam with energies in the range of MeV to GeV. The betatron oscillations of these electrons produce a collimated X-ray beam with ultrashort pulses known as betatron radiation [6] with energies in the order of units or tens of keV. Experimental scaling shows that the X-ray photon yield can be scaled up to $10^{12}$ photons/shot using PW class driving laser [7].

The driving laser of the Gammatron beamlines is the state-of-the-art HAPLS laser that can deliver up to 30 J at a 10 Hz repetition rate. The schematic of the Gammatron beamline is shown in Fig. 1 (a). From the technical point of view, the Gammatron beamline consists of four parts: (1) the laser focusing optics, for which up to several meters long focal length might be necessary, (2) the target - gas jet or gas cell, (3) several diagnostics necessary to optimize the electron and X-ray photon flux, and (4) the focusing and optimization of the X-ray and electron beams: X-ray mirror [8], magnetic lenses, filters, etc. The typical gases used as a target are pure He, He with the admixture of Nitrogen [9, 10], gas clusters, or dry air [11]. Prior to the operation, the gas jets are characterized using a high-resolution high-sensitivity interferometer [12,13].

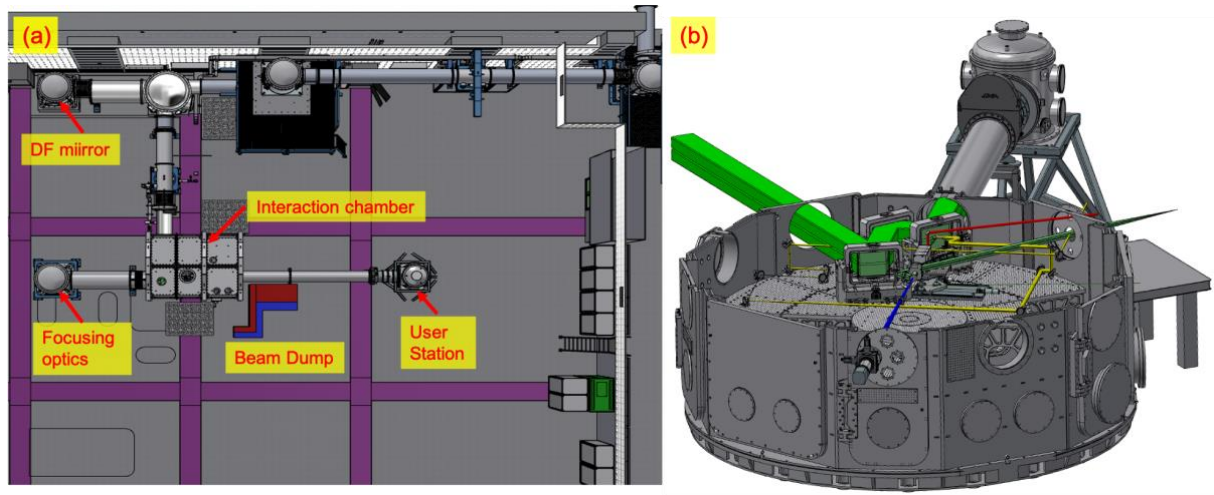

Fig. 1) a) Schematic of Gammatron beamline in the E2 experimental hall, b) Schematic of Betatron X-ray source inside P3 chamber in the experimental hall E3 at ELI Beamlines.

The laser-driven light sources operated within the Gammatron beamline, inherently produce femtosecond pulses of X-rays with a perfect temporal synchronization with the driving laser. Therefore, they offer a versatile tool in studying various time-resolved experiments in ultrafast science. In addition, with the tunable photon energy and an intrinsic synchronization to the driving IR laser, these sources are extremely useful for time-resolved structural analysis of matter with atomic resolution. Some of the applications of Gammatron beamlines are X-ray phase-contrast imaging [14], time-resolved X-ray absorption spectroscopy [15], X-ray diffraction, high-resolution radiography, and industrial imaging.

3. Betatron X-ray source for Plasma Physics Platform (P3)

The Plasma Physics Platform (P3) [16], located in Experimental Hall E3, is a versatile facility dedicated to fundamental research in laser plasma. The P3 infrastructure is designed to handle multiple synchronized laser beams (including the 10 PW, two high rep. rate 1 PW lasers, and a kilo-Joule uncompressed beam (long pulse ~ up to several ns)) in the same interaction chamber, covering a wide range of pulse lengths and energies [17]. The P3 infrastructure will focus on a wide range of experiments on laser-plasma physics including the ultra-intense laser-matter interaction, warm dense matter physics, HED physics [18, 19], adiabatic and radiative shock waves [20, 21], laboratory astrophysics, and advanced plasma physics experiments. Such experiments are key in benchmarking the astrophysical models and simulation codes [22].

The interaction of intense laser with matter produces very dense plasma. Probing such plasma requires a very bright hard X-ray source to penetrate it and to overcome the plasma self-emission. To overcome this, we are developing an ultra-short source of high-energy electron and betatron X-ray source in the P3 platform [5], the experimental scheme is shown in Fig 1 (b). The betatron source offers an ultrashort, collimated X-ray beam with a sub-radian divergence and micron source size. The ultrashort nature is making the source ideal backlighting tool to study time-resolved ultrafast processes in plasma physics experiments and the small source size allows performing phase-contrast imaging of fast-moving objects like radiative shock waves or jets.

## 4. Gas jet targets characterization station

We have developed a new optical probing technique for density characterization of gas jets that are used as LPA targets. We employ a novel optical configuration using both: two probe passes through the gas target and relay-imaging of the gas object between individual passes. The double pass of the probe beam through the phase object provides a two-fold increase of the accumulated phase.

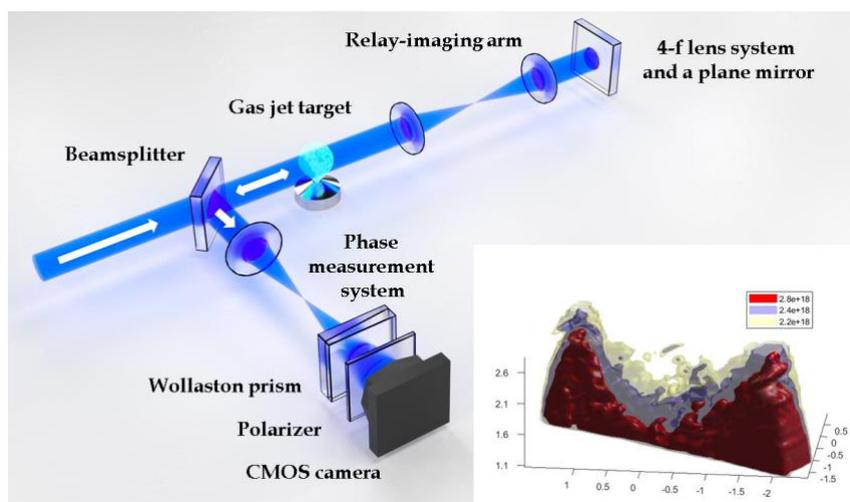

Fig. 2) Sketch of the 2-pass relay imaging Wollaston prism interferometer, Inset) Tomographic density reconstruction of a non-symmetric gas jet.

The optical configuration consists of a two-pass probing with a relay imaging object arm and a Wollaston prism interferometer. The relay imaging arm uses a combination of 4-f lens system of two achromatic doublet lenses and a plane mirror. This relay-imaging configuration preserves spatial information of the object thus allowing the diagnostics of gas jets with high sensitivity and high spatial resolution. We employed this interferometric setup for density characterization of gas jets with various spatial distributions: rotationally symmetric & non-rotationally symmetric, with tailored gas density distribution (pressure ramp and shock front). This technique can be further upgraded to a higher number of passes by introducing polarization switching of the probing laser beam 23 that would enhance the sensitivity further.

**Acknowledgments:** This research was supported by the project Advanced research (ADONIS) (CZ.02.1.01/0.0/0.0/16_019/0000789) and by the project High Field Initiative (HiFI) (CZ.02.1.01/0.0/0.0/15_003/0000449), both from European Regional Development Fund. The results of Project LQ1606 were obtained with the financial support of the Ministry of Education, Youth, and Sports as part of targeted support from the National Programme of Sustainability II.